\documentclass[12pt]{article}
\newcommand{\sect}[1]{\setcounter{equation}{0}\section{#1}}

\def\[{{[}}
\def\CD{{\cal D}}

\def\CP{{\cal P}}
\def\CL{{\cal L}}

\def\LL{{\mathbb L}}
\def\PP{{\mathbb P}}
\def\top{\hspace{-6mm}}

\def\1{\mathbb I}
\newcommand{\bea}{\begin{eqnarray}}
\newcommand{\eea}{\end{eqnarray}}

\newcommand{\nn}{\nonumber}

\newcommand{\pa}{\partial}
\newcommand{\p}[1]{(\ref{#1})}

\def\fnote#1#2{\begingroup\def\thefootnote{#1}\footnote{#2}\addtocounter
{footnote}{-1}\endgroup}

\usepackage{amsbsy,amssymb,latexsym}

\begin{document}
\begin{flushright}
KEK-TH-879 \\
OU-HET 437 \\
hep-th/0305215
\end{flushright}
\vspace{20mm}

\begin{center}
{\bf\Large
Uniqueness of M-theory PP-Wave Background \\
with Extra Supersymmetries
}

\vspace{20mm}
Nobuyoshi Ohta
\fnote{$\dagger$}{\texttt{ohta@phys.sci.osaka-u.ac.jp}}
and
Makoto Sakaguchi
\fnote{$^\star$}{\texttt{Makoto.Sakaguchi@kek.jp}}
\vspace{10mm}

\textit{
${^\dagger}$
Department of Physics, Osaka University, Toyonaka, Osaka 560-0043, JAPAN}\\
\vspace{3mm}
${}^\star$
\textit{Theory Division,
           High Energy Accelerator Research Organization (KEK)\\
           1-1 Oho, Tsukuba, Ibaraki, 305-0801, JAPAN}
\end{center}

\vspace{25mm}

\begin{abstract}

We examine Killing spinor equations of the general eleven-dimensional
pp-wave backgrounds, which contain a scalar $H(x^m,x^-)$ in the metric
and a three-form $\xi(x^m,x^-)$ in the flux. Considering non-harmonic
extra Killing spinors, we show that if the backgrounds admit at least
one extra Killing spinor in addition to the standard 16 Killing spinors,
they can be reduced to the form with $H=A_{mn}(x^-)x^mx^n$ and $\xi(x^-)$
modulo coordinate transformations. We further examine the cases in which
the extra Killing spinor is characterized by a set of Cartan matrices.
The super-isometry algebras of the resulting backgrounds are also derived.

\end{abstract}
\newpage

\sect{Introduction}

Supergravity solutions have played a central role in the study
of M-theory and superstring theories. Among them, pp-wave backgrounds
have attracted great interests recently.
In~\cite{Metsaev:Type IIB}, the Green-Schwarz superstring on the
maximally supersymmetric pp-wave background~\cite{BFHP:A new maximally}
was shown to be exactly solvable in the light-cone gauge,
and full string spectrum has been obtained. In~\cite{BMN}, considering
the large $N$ limit corresponding to the Penrose limit~\cite{Penrose},
AdS/CFT correspondence has been examined beyond the supergravity level,
and a matrix theory on the M-theory pp-wave background was proposed.

Eleven-dimensional pp-wave backgrounds~\cite{Hull:Exact pp wave}
\begin{eqnarray}
ds^2=2dx^+dx^-+H(x^-,x^m)(dx^-)^2+(dx^m)^2,~~~
F=dx^-\wedge \xi(x^-,x^m),
\label{pp:M}
\end{eqnarray}
are solutions of the eleven-dimensional supergravity theory when
\begin{eqnarray}
\triangle H=
-\frac{1}{3!}\xi_{lmn}\xi^{lmn},
\end{eqnarray}
where $\xi$ is a three-form on transverse $\mathbb E^9$ spanned by $x^m$.
These backgrounds admit at least sixteen standard Killing spinors.
At a special point in the moduli space, the background turns out to be
the Kowalski-Glikman (KG) solution~\cite{Kowalski-Glikman:vacuum}
which admits sixteen extra Killing spinors in addition to the sixteen
standard Killing spinors, and thus maximal thirty-two
supersymmetries~\cite{Chrusciel:The isometry}. The super-isometry algebra
of the KG solution has been obtained in~\cite{Figueroa-O'Farrill:Homogeneous}.
KG solution was shown to be obtained~\cite{BFCP:Penrose limits}
as a Penrose limit of AdS$_{4/7}\times S^{7/4}$ which is the near-horizon
limit of the M2/5-brane background. The super-isometry algebra of KG
solution was shown~\cite{HKS:Super-pp-wave} to be an In\"on\"u-Wigner (IW)
contraction of the super-isometry algebra of AdS$_{4/7}\times S^{7/4}$.
In addition to the cases with sixteen and thirty-two supersymmetries,
it has been shown that there exist pp-wave backgrounds with
18, 20, 22, 24, 26 supersymmetries~\cite{
CLP:Penrose,
Michelson:Twisted,
CLP:M-theory pp-waves,
GH:pp-waves in 11-dimensions,
Michelson:A pp-wave}.
Apart from special cases~\cite{
CLP:M-theory pp-waves,
LVP:Penrose,
Singh:M5-brane}
their AdS counterparts or brane intersections have not been understood well.

For the type-IIB supergravity theory, the maximally supersymmetric
pp-wave background and the super-isometry algebra have been obtained
in~\cite{BFHP:A new maximally}.
The super-isometry algebra was shown~\cite{HKS:IIB} to be derived
from that of AdS$_5\times S^5$ as an IW contraction.
It has been shown that there also exist non-maximally supersymmetric
pp-wave backgrounds~\cite{CLP:Penrose,
CLP:M-theory pp-waves,
CHKW:Penrose limit of RG fixed points,
BJLM:penrose limits deformed pp-waves,
BR:Supergravity,
GPS:penrose limit and RG flow}.
For the type-IIA supergravity theory, the maximally supersymmetric
pp-wave background does not exist \cite{FP;Maximally}.
The non-maximally cases were found in~\cite{Michelson:Twisted,
CLP:M-theory pp-waves,
Michelson:A pp-wave,
BR:Supergravity,
SY:IIA}.
For the lower dimensions, the maximally supersymmetric pp-wave backgrounds
were found in~\cite{Meessen:A small note} for five- and six-dimensions,
and in~\cite{Kowalski-Glikman:Positive} for four-dimensions.
The relation among these lower dimensional backgrounds
were discussed in~\cite{LMO:On d=4 5 6}.

In this paper, we examine the M-theory pp-wave background~(\ref{pp:M})
in the light of supersymmetries. Supersymmetries are determined by the
Killing spinor equation $\CD_M\varepsilon=0$ of the background.
For the maximally supersymmetric case \cite{FP;Maximally},
this condition reduces locally to the vanishing condition of
the curvature of the supercovariant connection $\CD_M$. Because the
Killing spinor $\varepsilon$ contains trivial entries for the non-maximally
case, it is not enough to examine the vanishing curvature condition of the
supercovariant connection $\CD_M$ in order to derive Killing spinors and
determine how many supersymmetries or Killing spinors are preserved.
We show that the M-theory pp-wave background~(\ref{pp:M}) is highly
restricted if there is at least one extra Killing spinor.
In particular, we find that the condition for the existence of at least one
extra Killing spinor restricts $H$ and $\xi$ in (\ref{pp:M}) to be
$A_{mn}(x^-)x^mx^n$ and $\xi(x^-)$, respectively. In addition, for the case
that the Killing spinor is characterized by a set of projectors only, $H$
and $\xi$ are found to reduce to $A_{m}(x^m)^2$ and $\xi$, respectively.
The resulting background is nothing but the pp-wave background assumed
in the literature. Thus, our consideration clarifies why and when
it is appropriate to consider the form of the pp-waves simply assumed in
the literature. Moreover, our consideration makes it clear how to construct
$x^-$-dependent (time-dependent) pp-wave backgrounds. We show in appendix B
that the time-dependent pp-wave background related to the anti-Mach
type background \cite{BMO} by a coordinate transformation \cite{HPW}
is characterized by the Killing spinor equation which is not expandable
with respect to projectors only.

This paper is organized as follows.
In the next section, we prove a uniqueness theorem which states that
$H(x^-,x^m)$ and $\xi(x^-,x^m)$ can be reduced to $A_{mn}(x^-)x^mx^n$
and a three-form $\xi(x^-)$, respectively, modulo coordinate transformations,
provided that the background admits at least one extra Killing spinor
in addition to the standard sixteen spinors. In section 3, we examine the
cases in which the extra Killing spinor is characterized by a set of Cartan
matrices. The super-isometry algebra of the resulting background is given
in section 4. The last section is devoted to a summary and discussions.

\sect{Uniqueness}
The general pp-wave background we consider in this section is
\begin{eqnarray}
ds^2&=&
 2dx^+dx^- +H(x^m,x^-)(dx^-)^2+(dx^m)^2,~~~
F=dx^-\wedge \xi(x^m,x^-)
\end{eqnarray}
where both the scalar $H$ and three-form $\xi$ on $\mathbb E^9$
spanned by $x^m$, are functions of $x^-$ and $x^m$.
This is a supergravity background when
\begin{eqnarray}
\triangle H=
-\frac{1}{3!}\xi_{lmn}\xi^{lmn},
\label{eom}
\end{eqnarray}
where $\triangle$ is the Laplacian on $\mathbb E^9$.
The frame one-forms defined by $ds^2=2e^+e^-+\delta_{mn}e^me^n$ are
\begin{eqnarray}
e^-=dx^-,~~~
e^+=dx^++\frac{1}{2}H(x^m,x^-)dx^-,~~~
e^m=dx^m
\end{eqnarray}
and thus the spin connection is given by
\begin{eqnarray}
w^+{}_m=\frac{1}{2}\partial_m H dx^-.
\end{eqnarray}
Killing spinor equations for general M-theory backgrounds
\begin{eqnarray}
&&\CD_M\varepsilon=(\nabla_M-\Omega_M)\varepsilon=0,
\\&&\nabla_M=\partial_M+\frac{1}{4}w_M^{ab}\Gamma_{ab},~~~
\Omega_M=\frac{1}{288}F_{PQRS}(\Gamma^{PQRS}{}_M+8\Gamma^{PQR}\delta^S_M),
\end{eqnarray}
boil down on this pp-wave background to
\begin{eqnarray}
\partial_+\varepsilon=0,~~~
\partial_-\varepsilon-\frac{1}{4}\partial_mH\Gamma^m\Gamma_+\varepsilon
 =\Omega_-\varepsilon,~~~
\partial_m\varepsilon=\Omega_m\varepsilon,
\label{KSEs}
\end{eqnarray}
where
\begin{eqnarray}
\Omega_m=\frac{1}{24}(\Gamma_m\Theta+3\Theta\Gamma_m)\Gamma_+,~~~
\Omega_-=-\frac{1}{12}\Theta(\Gamma_+\Gamma_-+1),~~~
\Theta=\frac{1}{3!}\xi_{lmn}\Gamma^{lmn}.
\end{eqnarray}
It is convenient to introduce nine-dimensional gamma-matrices $\gamma_m$
$\in Spin(9)$ by
\begin{eqnarray}
&&\Gamma_m=\gamma_m\otimes \sigma_3,~~~
\Gamma_\pm=\1_{16}\otimes\sigma_\pm,~~~
\sigma_\pm=\frac{1}{\sqrt{2}}(\sigma_1\pm i\sigma_2).
\label{gamma}
\end{eqnarray}
By the light-cone projection operator defined as
$\CP_\pm=\frac{1}{2}\Gamma_\pm\Gamma_\mp$,
Killing spinor $\varepsilon$ decomposes into
\begin{eqnarray}
\varepsilon=\left(
  \begin{array}{c}
   \varepsilon_+    \\
   \varepsilon_-    \\
  \end{array}
\right),~~~~
\CP_+\varepsilon=\left(
  \begin{array}{c}
   \varepsilon_+    \\
   0    \\
  \end{array}
\right),~~~
\CP_-\varepsilon=\left(
  \begin{array}{c}
   0    \\
   \varepsilon_-    \\
  \end{array}
\right),
\end{eqnarray}
where $\varepsilon_+$ is called the \textit{standard} Killing spinor
which exists for general pp-wave backgrounds, while $\varepsilon_-$ is
the \textit{extra} Killing spinor. In terms of $\varepsilon_\pm$,
Killing spinor equations~(\ref{KSEs}) are expressed as
\begin{eqnarray}
&&\partial_+\varepsilon_+=0,\label{1}\\
&&\partial_-\varepsilon_+-\frac{\sqrt{2}}{4}\partial_mH\gamma^m\varepsilon_-
 =-\frac{1}{4}\theta\varepsilon_+,
\label{2}\\
&&\partial_m\varepsilon_+=\frac{\sqrt{2}}{24}(\gamma_m\theta+3\theta\gamma_m)
\varepsilon_-,\label{3}\\
&&\partial_+\varepsilon_-=0 \label{4}\\
&&\partial_-\varepsilon_-=+\frac{1}{12}\theta
\varepsilon_-,\label{5}\\
&&\partial_m\varepsilon_-=0,\label{6}
\end{eqnarray}
where $
\theta=\frac{1}{3!}\xi_{lmn}\gamma^{lmn}$.
In the following,
we examine these equations and derive conditions on
$H$ and $\xi$, providing that
there exist at least one extra Killing spinor.

For non-maximally supersymmetric pp-wave backgrounds, $\varepsilon_-$
contains trivial entries. The projection operator $\PP_I$ of $\varepsilon_-$
into $I$-th non-trivial entry can be constructed as
\begin{eqnarray}
\PP_I=
diag (0,...,0,\stackrel{I}{1},0,...,0).
\end{eqnarray}
For the background which admits $N$ extra Killing spinors,
the projection is
\begin{eqnarray}
\PP=\sum_{I=I_1,...,I_N}\PP_I,~~~
\varepsilon_-=\PP\varepsilon_-.
\end{eqnarray}

It follows from~(\ref{4}) and (\ref{6}) that $\varepsilon_-$ is
independent of $x^+$ and $x^m$. This implies that $\theta \varepsilon_-$
depends only on $x^-$ from~(\ref{5}). In other words, $\theta\PP$ depends
only on $x^-$. {}From~(\ref{1}), we also see that $\varepsilon_+$ is
independent of $x^+$. Acting $\gamma^m$ on~(\ref{3}), one finds that
\begin{eqnarray}
\gamma^m\partial_m\varepsilon_+=0
\label{d}
\end{eqnarray}
because $\gamma^m(\gamma_m\theta+3\theta\gamma_m)=0$ for M-theory
pp-wave backgrounds.\footnote{Remember that
$\theta=\frac{1}{3!}\xi_{lmn}\gamma^{lmn}$.}
Further acting $\gamma^n\partial_n$ on this equation,
we find that $\varepsilon_+$ satisfies the Laplace equation
\begin{eqnarray}
\partial^m\partial_m\varepsilon_+=0.
\label{dd}
\end{eqnarray}
Consequently $\varepsilon_+$ must be linear in $x^m$ at most,
up to a harmonic function.
In this paper, we concentrate on
the non-harmonic function part.\footnote{
It is interesting to include the harmonic function
into the study
and examine the resulting background.
}
We can thus write $\varepsilon_+$ as
\begin{eqnarray}
\varepsilon_+=\varepsilon_0(x^-)+\varepsilon_m(x^-)x^m
\end{eqnarray}
where $\varepsilon_0$ and $\varepsilon_m$ are functions of $x^-$ only
and (\ref{3}) becomes
\begin{eqnarray}
\varepsilon_m=\frac{\sqrt{2}}{24}(\gamma_m\theta+3\theta\gamma_m)\varepsilon_-.
\label{varepsilon_m}
\end{eqnarray}
Since $\varepsilon_m$ and $\theta\PP$ depend only on $x^-$, so does
$\theta\gamma_m\PP$ for all $m$. Because $\PP$ is made of products of
$\gamma_m$ as will be seen in subsections 2.1 and 2.2,
$\gamma_m\PP_I=\PP_J\gamma_m$ for generic $I\neq J$.
If both $\PP_I$ and $\PP_J$ are contained in $\PP$, $\gamma_m\PP=\PP\gamma_m$,
whereas $\gamma_m\PP=\overline{\PP}\gamma_m$ if $\PP_I$ is contained in
$\PP$ but $\PP_J$ is not, where $\overline{\PP}$ is the projection operator
complementary to $\PP$ satisfying $\PP+\overline{\PP}=\1_{16}$.
Thus the term $\theta\gamma_m\PP$ is equal either to $\theta\PP\gamma_m$
for a certain set of $m$ or to $\theta\overline{\PP}\gamma_m$ for the rest
of $m$. No new condition arises in the former case, whereas
in the latter case $\theta\overline{\PP}$ must depend only on $x^-$.
As a result, from the fact that both $\theta\PP$ and $\theta\overline{\PP}$
depend only on $x^-$, which are derived below (2.18) and above,
respectively, we can conclude that $\theta$ depends only on $x^-$.
{}From~(\ref{2}), $\partial_mH$ must be linear in $x^m$ at most,
and thus we can write $H$ as $H=f(x^-)+g_m(x^-)x^m+A_{mn}(x^-)x^mx^n$,
where $f,~g_m$ and $A_{mn}$ are functions of $x^-$ only.

To summarize up to this point, we have shown that M-theory pp-wave
backgrounds with extra supersymmetries can be reduced to the form
\begin{eqnarray}
ds^2&=&
2dx^+dx^- +\Big(f(x^-)+g_m(x^-)x^m+A_{mn}(x^-)x^mx^n\Big)(dx^-)^2+(dx^m)^2,
\nonumber\\
F&=&dx^-\wedge \xi(x^-).
\end{eqnarray}

We now argue that $f(x^-)$ and $g_m(x^-)$ can be absorbed by a coordinate
redefinition
\begin{eqnarray}
x^+ = y^+ - F(x^-) - G_m(x^-)y^m,~~~
x^m = y^m - H^m(x^-).
\label{redefinition}
\end{eqnarray}
The line element then becomes
\begin{eqnarray}
ds^2&=& 2dx^- \Big[ dy^+ +dx^- \{ -\pa F+\frac{1}{2}f-g_mH^m
+\frac{1}{2}A_{mn}H^mH^n + \frac{1}{2}(\pa H^m)^2 \}
\nonumber\\&&~~~~
+ dx^- y^m \{ -\pa G_m +\frac{1}{2}g_m -A_{mn}H^n \}
  +dy^m \{ -G_m -\pa H_m \} + \frac{1}{2}A_{mn} y^m y^n dx^- \Big]
\nonumber  \\&&~~~~
+(dy^m)^2.
\end{eqnarray}
We see that if we choose $F, G_m$ and $H^m$ such that
\begin{eqnarray}
&&-\pa F+\frac{1}{2}f-g_mH^m+\frac{1}{2}A_{mn}H^mH^n
 + \frac{1}{2}(\pa H^m)^2=0,\nn \\
&& -\pa G_m +\frac{1}{2}g_m -A_{mn}H^n =0,\\
&&-G_m -\pa H_m=0, \nn
\end{eqnarray}
the line element reduces to
\begin{eqnarray}
ds^2&=& 2dy^+dx^- + A_{mn} y^m y^n (dx^-)^2 +(dy^m)^2.
\end{eqnarray}
The transformation~(\ref{redefinition}) does not affect $F=dx^-\wedge \xi$.

In summary, we have shown that M-theory pp-wave backgrounds
which admit extra Killing spinors\footnote{
It should be noted again that we are considering non-harmonic extra
Killing spinors $\varepsilon_+$.
} can be reduced to the form
\begin{eqnarray}
ds^2=
 2dx^+dx^- +A_{mn}(x^-)x^mx^n(dx^-)^2+(dx^m)^2,~~~
F=dx^-\wedge \xi(x^-),
\label{pp-}
\end{eqnarray}
modulo coordinate transformations.

$A_{mn}(x^-)$ and $\xi(x^-)$ are restricted by the condition (\ref{2}).
On the background (\ref{pp-}), (\ref{2}) becomes
\begin{eqnarray}
\partial_-\varepsilon_m-\frac{\sqrt{2}}{2}A_{mn}\gamma^n\varepsilon_-
 =-\frac{1}{4}\theta\varepsilon_m,
\end{eqnarray}
which, together with (\ref{varepsilon_m}), implies that
\begin{eqnarray}
[
12 \partial_-U_{(m)}
+U_{(m)}\theta
+3\theta_{(m)} U_{(m)}
+V_{(m)}
-12^2A_{m} ]\varepsilon_-\equiv D_{(m)}\varepsilon_-=0,
\label{KSE:before}
\end{eqnarray}
where
\begin{eqnarray}
U_{(m)}\equiv \theta+3\theta_{(m)},~~~
\gamma_m\theta_{(m)}\equiv \theta\gamma_m,~~~
V_{(m)}\equiv -12^2\sum_{n=1}^9A_{mn}\gamma^{mn},~~~
A_m\equiv A_{mm}.
\end{eqnarray}
We examine this condition in the next section for the cases in which
$D_{(m)}$ is expanded solely in terms of mutually commuting projectors.

\sect{$D_{(m)}$ expandable in mutually commuting projectors}

For the existence of extra Killing spinors, $D_{(m)}$ must be a linear
combination of projection operators. Here we restrict our study to the
case in which $D_{(m)}$ is expanded \textit{solely} in terms of mutually
commuting projectors.\footnote{More general cases are studied in \cite{BMO}.
We thank the authors for explanation of their work.}
Projection operators are composed of Cartan matrices, $H_I$, as
\begin{eqnarray}
P_I=\frac{1}{2}(\1+H_I).
\end{eqnarray}
There are infinitely many Cartan matrices. Among them, we consider the
simplest case in which Cartan matrices are monomials of gamma-matrices,
so that $H_I=\gamma^{[N]}$ where $\gamma^{[N]}$ is
an $N$-th antisymmetrized product of gamma matrices.
Other Cartan matrices are obtained by similarity transformations
from the simplest case. We will discuss some of them in subsection 3.2.

\subsection{The simplest case}

In this case, it is sufficient to consider $N_I=1,2,3,4$, because
$\gamma^{[N]}$ is related to $\gamma^{[9-N]}$ using $\gamma^{1...9}=\1$.
Noting that for both $M$ and $N$ odd
\begin{eqnarray}
\[\gamma^{m_1...m_M},\gamma^{n_1...n_N}]&=&
2\gamma^{m_1...m_Mn_1...n_N}
\\&&\hspace{-10mm}
+2\sum_{i=1}^{[\frac{min(M,N)}{2}]}
\Big[(2i)!
\left(\hspace{-2mm}
  \begin{array}{c}
    M   \\
    2i   \\
  \end{array}
\hspace{-2mm}\right)\left(\hspace{-2mm}
  \begin{array}{c}
    N   \\
    2i   \\
  \end{array}
\hspace{-2mm}\right)
\delta^{m_Mn_1}\cdots \delta^{m_{M-2i+1}n_{2i}}
\gamma^{m_1...m_{M-2i}n_{2i+1}...n_N}
\Big],
\nonumber
\end{eqnarray}
and for either $N$ or $M$ even
\begin{eqnarray}
\[\gamma^{m_1...m_M},\gamma^{n_1...n_N}]&=&2\sum_{i=1}^{[\frac{min(M,N)+1}{2}]}
\Big[
(2i-1)!\left(\hspace{-2mm}
  \begin{array}{c}
    M   \\
    2i-1   \\
  \end{array}
\hspace{-2mm}\right)\left(\hspace{-2mm}
  \begin{array}{c}
    N   \\
    2i-1   \\
  \end{array}
\hspace{-2mm}\right)
\\&&\hspace{20mm}
\times
\delta^{m_Mn_1}\cdots \delta^{m_{M-2i+2}n_{2i-1}}
\gamma^{m_1...m_{M-2i+1}n_{2i}...n_N}
\Big]
,\nonumber
\end{eqnarray}
where the appropriate antisymmetrization of the indices is understood on
the right hand side, one finds that mutually commuting matrices $\gamma^{[N]}$
must share a definite number of indices. We indicate the number of common
indices shared among two of matrices below.
\begin{eqnarray}
  \begin{array}{c|cccc}
&\gamma^{n}&\gamma^{n_1n_2}&\gamma^{n_1n_2n_3}&\gamma^{n_1\cdots n_4}\\
\hline
\gamma^{m}& 1   & 0   & 1   & 0   \\
\gamma^{m_1m_2}& 0   &0,2    &0,2    &0,2    \\
\gamma^{m_1m_2m_3}&1    & 0,2   &1,3    &0,2    \\
\gamma^{m_1\cdots m_4}& 0   &0,2    &0,2    &0,2,4    \\
  \end{array}
\end{eqnarray}
We find two sets of mutually commuting fifteen matrices;\footnote{
The first set (\ref{matrices-1}) can be related to the K\"ahler
form $J$ of a Calabi-Yau four-fold with SU(4) holonomy. The K\"ahler form
$J$ is covariantly constant $dJ=0$. The terms in the second line are the
constituents of $J$, and those in the fourth, third and first lines are the
constituents of $J\wedge J$, $J\wedge J\wedge J$ and $J\wedge J\wedge J
\wedge J$, respectively. On the other hand, the second set (\ref{matrices-2})
can be related to the associative three-form $\phi$ of $d=7$ Riemannian
manifold with $G_2$ holonomy, which is covariantly constant $d\phi=0$.
The terms in the second line are the constituents of $\phi$, and those
in the third line are the constituents of the seven-dimensional Hodge dual
of $\phi$, $*_7\phi$. The first line is $\phi\wedge *_7\phi$.} one is
\begin{eqnarray}
&&
\gamma^9
\nonumber\\&&
\gamma^{12},
\gamma^{34},
\gamma^{56},
\gamma^{78}
\nonumber\\&&
\gamma^{129},
\gamma^{349},
\gamma^{569},
\gamma^{789}
\nonumber\\&&
\gamma^{1234},
\gamma^{3456},
\gamma^{5678},
\gamma^{1256},
\gamma^{1278},
\gamma^{3478}
\label{matrices-1}
\end{eqnarray}
and the other is
\begin{eqnarray}
&&
\gamma^{89}
\nonumber\\&&
\gamma^{123},
\gamma^{145},
\gamma^{167},
\gamma^{246},
\gamma^{257},
\gamma^{347},
\gamma^{356},
\nonumber\\&&
\gamma^{4567},
\gamma^{2367},
\gamma^{2345},
\gamma^{1357},
\gamma^{1346},
\gamma^{1256},
\gamma^{1247}.
\label{matrices-2}
\end{eqnarray}
The former is related to the latter
by the similarity transformation
\begin{eqnarray}
H_I\to SH_IS^{-1},~~~
S=e^{\frac{\pi}{4}\gamma^{246}}
\end{eqnarray}
and the renaming of indices, $(1,2,...,8,9)\to (2,3,...,9,1)$.
Though both cases turn out to lead to the same result, the reasoning is
considerably different and we present the analysis of the former case here,
and the latter case is relegated to the appendix.

Now $D_{(m)}$ is constructed from (\ref{matrices-1}), and thus
$\theta=\frac{1}{3!}\xi_{lmn}\gamma^{lmn}$ takes the form
\begin{eqnarray}
\theta=
 a_1\gamma^{129}
 +a_2\gamma^{349}
 +a_3\gamma^{569}
 +a_4\gamma^{789}
 \label{theta1}.
\end{eqnarray}
$A_{mn}$ is restricted to be non-vanishing only when $m=n$ and
$(m,n)=(1,2),~(3,4),~(5,6),~(7,8)$. Because terms in (\ref{theta1})
commute with each other, (\ref{KSE:before}) reduces to
\begin{eqnarray}
[12\partial_-U_{(m)}+U_{(m)}^2+V_{(m)}-12^2A_m]\varepsilon_-=0.
\label{KSE}
\end{eqnarray}
$U_{(m)}$ and $V_{(m)}$ for~\p{matrices-1} is expressed as
\begin{eqnarray}
U_{(m)}&=&
\alpha_1^m\gamma^{129}
+\alpha^m_2\gamma^{349}
+\alpha^m_3\gamma^{569}
+\alpha^m_4\gamma^{789},
\nonumber\\
V_{(m)}&=&
 \mu_1^m\gamma^{12}
+\mu_2^m\gamma^{34}
+\mu_3^m\gamma^{56}
+\mu_4^m\gamma^{78},
\label{U1}
\end{eqnarray}
where
\begin{eqnarray}
\alpha_1&=&(4a_1,4a_1,-2a_1,-2a_1,-2a_1,-2a_1,-2a_1,-2a_1,4a_1),\nn\\
\alpha_2&=&(-2a_2,-2a_2,4a_2,4a_2,-2a_2,-2a_2,-2a_2,-2a_2,4a_2),\nn\\
\alpha_3&=&(-2a_3,-2a_3,-2a_3,-2a_3,4a_3,4a_3,-2a_3,-2a_3,4a_3),\nn\\
\alpha_4&=&(-2a_4,-2a_4,-2a_4,-2a_4,-2a_4,-2a_4,4a_4,4a_4,4a_4),\nn\\
\mu_i^{2i-1}&=&-\mu_i^{2i}=-12^2A_{2i-1~2i},~~~~i=1,2,3,4.
\label{alpha}
\end{eqnarray}

In this case, all the matrices (\ref{matrices-1}) can be
constructed as products of four matrices, $\gamma^{129}$, $\gamma^{349}$,
$\gamma^{569}$ and $\gamma^{789}$. From these matrices, we make
four rank-8 projection operators
\begin{eqnarray}
&&P_1=\frac{1}{2}(\1+i\gamma^{129}),~~~
P_2=\frac{1}{2}(\1+i\gamma^{349}),
\nonumber\\&&
P_3=\frac{1}{2}(\1+i\gamma^{569}),~~~
P_4=\frac{1}{2}(\1+i\gamma^{789}),
\label{P1}
\end{eqnarray}
which satisfy
\begin{eqnarray}
P_A^2=P_A,~~~
P_AP_B=P_BP_A,~~~
A,B=1,2,3,4.
\end{eqnarray}
We rewrite (\ref{KSE}) in terms of these projection operators.
Because
\begin{eqnarray}
&&
\gamma^{129}=-i(2P_1-\1),~~~
\gamma^{349}=-i(2P_2-\1),~~~
\gamma^{569}=-i(2P_3-\1),~~~
\gamma^{789}=-i(2P_4-\1),
\nonumber\\&&
\gamma^{12}=-i(2P_2-\1)(2P_3-\1)(2P_4-\1),~~~
\gamma^{34}=-i(2P_1-\1)(2P_3-\1)(2P_4-\1),
\nonumber\\&&
\gamma^{56}=-i(2P_1-\1)(2P_2-\1)(2P_4-\1),~~~
\gamma^{78}=-i(2P_1-\1)(2P_2-\1)(2P_3-\1),
\end{eqnarray}
$U_{(m)}$ and $V_{(m)}$ in (\ref{U1}) can be written as
\begin{eqnarray}
U_{(m)}&=&
i(\alpha_1^m+\alpha^m_2+\alpha^m_3+\alpha^m_4)\1
-2i(\alpha_1^mP_1+\alpha^m_2P_2+\alpha^m_3P_3+\alpha^m_4P_4),\nn\\
V_{(m)}&=&
i(\mu_1^m+\mu_2^m+\mu_3^m+\mu_4^m)\1
-2i(\mu_2^m+\mu_3^m+\mu_4^m)P_1
-2i(\mu_1^m+\mu_2^m+\mu_3^m)P_2
\nonumber\\&&
-2i(\mu_1^m+\mu_2^m+\mu_4^m)P_3
-2i(\mu_1^m+\mu_2^m+\mu_3^m)P_4
+4i(\mu_1^m+\mu_4^m)P_2P_3
\nonumber\\&&
+4i(\mu_3^m+\mu_4^m)P_1P_2
+4i(\mu_1^m+\mu_2^m)P_3P_4
+4i(\mu_1^m+\mu_3^m)P_2P_4
\nonumber\\&&
+4i(\mu_2^m+\mu_4^m)P_1P_3
+4i(\mu_2^m+\mu_3^m)P_1P_4
-8i\mu_1^mP_2P_3P_4
-8i\mu_2^mP_1P_3P_4
\nonumber\\&&
-8i\mu_3^mP_1P_2P_4
-8i\mu_4^mP_1P_2P_3.
\end{eqnarray}
Substituting these into (\ref{KSE}) yields
\begin{eqnarray}
&\Bigl[&
\Big(12i\partial_-(
 \alpha^m_1
 +\alpha^m_2
 +\alpha^m_3
 +\alpha^m_4
 )
+i(\mu_1^m+\mu_2^m+\mu_3^m+\mu_4^m)
\nonumber\\&&~~~~~~~~
-(\alpha^m_1+\alpha^m_2+\alpha^m_3+\alpha^m_4
)^2
-12^2 A_m\Big)\1
\nonumber\\&&~~
+4\Big(
 -6 i\partial_-\alpha^m_1
 -\frac{i}{2}(\mu_2^m+\mu_3^m+\mu_4^m)
 +\alpha^m_1
 (\alpha^m_2
 + \alpha^m_3
 + \alpha^m_4)
 \Big) P_1
\nonumber\\&&~~
+4\Big(
 -6 i\partial_-\alpha^m_2
-\frac{i}{2}(\mu_1^m+\mu_2^m+\mu_3^m)
 + \alpha^m_2
 (\alpha^m_4
 + \alpha^m_1
 + \alpha^m_3 )
 \Big) P_2
\nonumber\\&&~~
+4\Big(
 -6i\partial_-\alpha^m_3
-\frac{i}{2}(\mu_1^m+\mu_2^m+\mu_4^m)
 + \alpha^m_3
 (\alpha^m_4
 + \alpha^m_2
 + \alpha^m_1 )
 \Big) P_3
\nonumber\\&&~~
+4\Big(
 -6 i\partial_-\alpha^m_4
 -\frac{i}{2}(\mu_1^m+\mu_2^m+\mu_3^m)
 +  \alpha^m_4
 (\alpha^m_1
 + \alpha^m_2
 + \alpha^m_3)
 \Big) P_4
\nonumber\\&&~~
+\Big(
 4i(\mu_3^m+\mu_4^m)
 -8 \alpha^m_1\alpha^m_2
\Big)P_1 P_2
+\Big(
 4i(\mu_2^m+\mu_4^m)
 -8 \alpha^m_1\alpha^m_3
\Big)P_1 P_3
\nonumber\\&&~~
+\Big(
 4i(\mu_2^m+\mu_3^m)
 -8 \alpha^m_1\alpha^m_4
\Big) P_1 P_4
+\Big(
 4i(\mu_1^m+\mu_4^m)
 -8 \alpha^m_2\alpha^m_3
\Big)P_2 P_3
\nonumber\\&&~~
+\Big(
 4i(\mu_1^m+\mu_3^m)
 -8 \alpha^m_2\alpha^m_4
\Big)P_2 P_4
+\Big(
 4i(\mu_1^m+\mu_2^m)
 -8 \alpha^m_3\alpha^m_4
\Big)P_3 P_4
\nonumber\\&&~~
-8i\mu_1^mP_2P_3P_4
-8i\mu_2^mP_1P_3P_4
-8i\mu_3^mP_1P_2P_4
-8i\mu_4^mP_1P_2P_3
\Bigr]\varepsilon_-=0.
\label{KSE1}
\end{eqnarray}
In order to see which Killing spinor survives, it is convenient to introduce
rank-1 projection operators of a 16-component spinor onto the $I$-th component:
\begin{eqnarray}
\PP_I=diag(0,\ldots,0,\stackrel{I}{1},0,\ldots,0).
\end{eqnarray}
The rank-8 projection operators $P_A$ can then be expressed in terms of
these rank-1 projection operators as
\begin{eqnarray}
&&P_1=\sum_{I=1,2,..,8}\PP_I,~~~
P_2=\sum_{I=1,2,3,4,9,10,11,12}\PP_I,\nonumber\\&&
P_3=\sum_{I=1,2,5,6,9,10,13,14}\PP_I,~~~
P_4=\sum_{I=1,3,5,7,9,11,13,15}\PP_I,
\label{PP1}
\end{eqnarray}
and $\sum_{I=1,..,16}\PP_I=\1_{16}$. In terms of these projection
operators, (\ref{KSE1}) becomes
\begin{eqnarray}
&
\Bigl[&
\Big(
12i\partial_-(
\alpha_1^m+\alpha_2^m+\alpha_3^m+\alpha_4^m
)
+i(\mu_1^m+\mu_2^m+\mu_3^m+\mu_4^m)
\nonumber\\&&~~~~~~~~
-(\alpha_1^m+\alpha_2^m+\alpha_3^m+\alpha_4^m)^2
-12^2 A_m
\Big)\1
\nonumber\\&&~~
+\Big(
-24i\partial_-(
\alpha_1^m+\alpha_2^m+\alpha_3^m+\alpha_4^m
)
-2i(\mu_1^m+\mu_2^m+\mu_3^m+\mu_4^m)
\Big)\PP_1
\nonumber\\&&~~
+\Big(
-24i\partial_-(
\alpha_1^m+\alpha_2^m+\alpha_3^m
)
-2i\mu_4^m
+4\alpha_4^m(\alpha_1^m
+\alpha_2^m
+\alpha_3^m)
\Big)\PP_2
\nonumber\\&&~~
+\Big(
-24i\partial_-(
\alpha_1^m+\alpha_2^m+\alpha_4^m
)
-2i\mu_3^m
+4\alpha_3^m(\alpha_1^m+\alpha_2^m+\alpha_4^m)
\Big)\PP_3
\nonumber\\&&~~
+\Big(
-24i\partial_-(\alpha_1^m+\alpha_2^m)
-2i(\mu_1^m+\mu_2^m)
+4(\alpha_1^m+\alpha_2^m)(\alpha_3^m+\alpha_4^m)
\Big)\PP_4
\nonumber\\&&~~
+\Big(
-24i\partial_-(\alpha_1^m+\alpha_3^m+\alpha_4^m)
-2i\mu_2^m
+4\alpha_2^m(\alpha_1^m+\alpha_3^m+\alpha_4^m)
\Big)\PP_5
\nonumber\\&&~~
+\Big(
-24i\partial_-(\alpha_1^m+\alpha_3^m)
-2i(\mu_1^m+\mu_3^m)
+4(\alpha_1^m+\alpha_3^m)(\alpha_2^m+\alpha_4^m)
\Big)\PP_6
\nonumber\\&&~~
+\Big(
-24i\partial_-(\alpha_1^m+\alpha_4^m)
-2i(\mu_1^m+\mu_4^m)
+4(\alpha_1^m+\alpha_4^m)(\alpha_2^m+\alpha_3^m)
\Big)\PP_7\nonumber\\&&~~
+\Big(
-24i\partial_- \alpha^m_1
-2i(\mu_2^m+\mu_3^m+\mu_4^m)
+4 \alpha^m_1(\alpha^m_2+ \alpha^m_3
+\alpha^m_4
)
\Big)\PP_8\nonumber\\&&~~
+\Big(
-24i\partial_-(
\alpha^m_2+\alpha^m_3+\alpha^m_4
)
-2i\mu_1^m
+4 \alpha^m_1( \alpha^m_2
+\alpha^m_3
+\alpha^m_4)
\Big)\PP_9\nonumber\\&&~~
+\Big(
-24i\partial_-(
\alpha^m_2+\alpha^m_3
)
-2i(\mu_2^m+\mu_3^m)
+4 (\alpha^m_1+\alpha^m_4)( \alpha^m_2+\alpha^m_3)
\Big)\PP_{10}\nonumber\\&&~~
+\Big(
-24i\partial_-(
\alpha^m_2+\alpha^m_4
)
-2i(\mu_2^m+\mu_4^m)
+4(\alpha^m_1+\alpha^m_3 )(\alpha^m_2+\alpha^m_4)
\Big)\PP_{11}\nonumber\\&&~~
+\Big(
-24i\partial_-\alpha^m_2
-2i(\mu_1^m+\mu_3^m+\mu_4^m)
+4 \alpha^m_2(  \alpha^m_1 +\alpha^m_3+\alpha^m_4)
\Big)\PP_{12}\nonumber\\&&~~
+\Big(
-24i\partial_-(\alpha^m_3+\alpha^m_4)
-2i(\mu_3^m+\mu_4^m)
+4(\alpha^m_1+\alpha^m_2 )(\alpha^m_3+ \alpha^m_4)
\Big)\PP_{13}\nonumber\\&&~~
+\Big(
-24i\partial_-\alpha^m_3
-2i(\mu_1^m+\mu_2^m+\mu_4^m)
+4\alpha^m_3(\alpha^m_1+\alpha^m_2+ \alpha^m_4)
\Big)\PP_{14}\\&&~~
+\Big(
-24i\partial_-\alpha^m_4
-2i(\mu_1^m+\mu_2^m+\mu_3^m)
+4\alpha^m_4(\alpha^m_1+ \alpha^m_2+\alpha^m_3 )
\Big)\PP_{15}
~~~
\Bigr]\varepsilon_-=0.
\nonumber
\end{eqnarray}
In order to have an extra Killing spinor $\PP_{16}\varepsilon_-$,
the coefficient of $\1$ must vanish so that
\begin{eqnarray}
&&12\partial_-(
\alpha_1^m+\alpha_2^m+\alpha_3^m+\alpha_4^m)
+(\mu_1^m+\mu_2^m+\mu_3^m+\mu_4^m)=0,\label{partial}\\
&&A_m=-\frac{1}{12^2}
(\alpha_1^m+\alpha_2^m+\alpha_3^m+\alpha_4^m)^2,
\label{A_m}
\end{eqnarray}
because the first line of the coefficient of $\1$ is imaginary while the
second line is real, and thus these two parts must vanish
separately.\footnote{The condition for another spinor, say
$\PP_{15}\varepsilon_-$ instead of $\PP_{16}\varepsilon_-$, to be an
extra Killing spinor is simply obtained from (\ref{partial}) and (\ref{A_m})
by the replacements $\alpha_4^m\to-\alpha_4^m$ and $\mu_i^m\to-\mu_i^m$,
$i=1,2,3$ using the relation $\sum_{I=1}^{16}\PP_I=\1$.
Thus, without loss of generality, we can take (\ref{partial}) and
(\ref{A_m}) as the condition for the existence of the extra Killing spinors.}
The first condition (\ref{partial}) leads to
\begin{eqnarray}
&&
  \begin{array}{l}
12\partial_-(\alpha_1^{2n-1}+\alpha_2^{2n-1}+\alpha_3^{2n-1}+\alpha_4^{2n-1})
 +\mu_n^{2n-1}=0,       \\
12\partial_-(\alpha_1^{2n}+\alpha_2^{2n}+\alpha_3^{2n}+\alpha_4^{2n})
 +\mu_n^{2n}=0,       \\
  \end{array}
  ~~~
n=1,2,3,4,\nonumber\\
&&~
12\partial_-(\alpha_1^{9}+\alpha_2^{9}+\alpha_3^{9}+\alpha_4^{9})=0,
\end{eqnarray}
from which we find that $\mu_i^m=0$. This means $A_{12}=A_{34}=A_{56}=A_{78}
=0$, and
\begin{eqnarray}
\partial_-(\alpha_1^m+\alpha_2^m+\alpha_3^m+\alpha_4^m)=0,
\label{partial'}
\end{eqnarray}
because $\alpha_i^{2n-1}=\alpha_i^{2n}$ and $\mu_n^{2n-1}=-\mu_n^{2n}$
as recognized from (\ref{alpha}).
Consequently $A_m$ must be independent of $x^-$, because the right hand
side of (\ref{A_m}) is independent of $x^-$ from eq.~(\ref{partial'}).
Noting that $\alpha_i^m$ is related to $a_i$ in (\ref{alpha}),
we find that eq.~(\ref{partial'}) leads to four differential equations
for $a_i$:
\begin{eqnarray}
&&
\partial_-(4a_1-2a_2-2a_3-2a_4)=0,~~~
\partial_-(-2a_1+4a_2-2a_3-2a_4)=0,\nonumber\\&&
\partial_-(-2a_1-2a_2+4a_3-2a_4)=0,~~~
\partial_-(-2a_1-2a_2-2a_3+4a_4)=0,
\end{eqnarray}
which imply
\begin{eqnarray}
\partial_-a_1=\partial_-a_2=\partial_-a_3=\partial_-a_4=0.
\end{eqnarray}
This means that $\xi$ is independent of $x^-$.
The extra Killing spinors are determined as a non-trivial solution of
\begin{eqnarray}
&&
\Bigl[
(
-(\alpha_1^m+\alpha_2^m+\alpha_3^m+\alpha_4^m)^2
-12^2 A_m
)\1_{16}
\nonumber\\&&~~
+4\alpha_4^m(\alpha_1^m
+\alpha_2^m
+\alpha_3^m)
(\PP_2+\PP_{15})
+4\alpha_3^m(\alpha_1^m+\alpha_2^m+\alpha_4^m)
(\PP_3+\PP_{14})
\nonumber\\&&~~
+4(\alpha_1^m+\alpha_2^m)(\alpha_3^m+\alpha_4^m)
(\PP_4+\PP_{13})
+4\alpha_2^m(\alpha_1^m+\alpha_3^m+\alpha_4^m)
(\PP_5+\PP_{12})
\nonumber\\&&~~
+4(\alpha_1^m+\alpha_3^m)(\alpha_2^m+\alpha_4^m)
(\PP_6+\PP_{11})
+4(\alpha_1^m+\alpha_4^m)(\alpha_2^m+\alpha_3^m)
(\PP_7+\PP_{10})\nonumber\\&&~~
+4 \alpha^m_1(\alpha^m_2+ \alpha^m_3
+\alpha^m_4)
(\PP_8+\PP_9)
~~~~
\Bigr]\varepsilon_-=0,
\end{eqnarray}
which reveals the two-fold degeneracy of the extra Killing
spinors. If (\ref{A_m}) is satisfied, $\PP_1\varepsilon_-$ and $\PP_{16}
\varepsilon_-$ are a pair of the extra Killing spinors and the background
admits 18 Killing spinors, 16 standard and 2 extra Killing spinors.
If, in addition, the coefficient of $(P_I+P_{17-I})$ is zero,
then $P_I\varepsilon_-$ and $P_{17-I}\varepsilon_-$
give another pair of extra Killing spinors.
Examining these conditions, one obtains pp-wave backgrounds which admit
18, 20, 22, 24, 32 Killing spinors \cite{GH:pp-waves in 11-dimensions}.

We find from (\ref{A_m}) and (\ref{alpha}) that
\begin{eqnarray}
\triangle H=\sum_{m=1}^92A_m=-a_1^2-a_2^2-a_3^2-a_4^2,
\end{eqnarray}
and from (\ref{theta1}) that
\begin{eqnarray}
-\frac{1}{3!}\xi_{lmn}\xi^{lmn}=-a_1^2-a_2^2-a_3^2-a_4^2.
\end{eqnarray}
Thus we see that the supergravity equation of motion
is automatically satisfied
for pp-wave backgrounds with extra supersymmetries
characterized by Cartan matrices (\ref{matrices-1}).
\bigskip

In summary, we have shown in this section and the appendix A
that M-theory pp-wave backgrounds which admit extra Killing spinors
characterized by Cartan matrices (\ref{matrices-1}) and (\ref{matrices-2})
can be reduced to the form
\begin{eqnarray}
ds^2=
 2dx^+dx^- +A_{m}x^mx^m(dx^-)^2+(dx^m)^2,~~~
F=dx^-\wedge \xi,
\label{pp}
\end{eqnarray}
modulo coordinate transformations, where $A_m$ and $\xi$ are constants.

In the next subsection, we will examine more general cases
that the Killing spinors are not characterized by monomial Cartan matrices,
and show that the background reduces to (\ref{pp}) again.
This suggests that the background reduces to (\ref{pp}) when the Killing
spinors are characterized by projectors only. Indeed we show in appendix B
that the Killing spinors on the time-dependent pp-wave background
related to the anti-Mach type background \cite{BMO}
are not characterized by projectors only.

\subsection{more general cases}

We consider the case in which some of Cartan matrices are not monomials.
As an example, let us consider the similarity transformation
\begin{eqnarray}
H_I\to SH_IS^{-1},~~~
S=e^{\theta\gamma^{123}}
\end{eqnarray}
which transforms
(\ref{matrices-1})
into
\begin{eqnarray}
&&
 a\gamma^9+b\gamma^{1239},~~
 \gamma^{12}~~,
 a\gamma^{34}+b\gamma^{124},~~
 \gamma^{56},~~
 \gamma^{78},
\nonumber\\&&
 a\gamma^{129}-b\gamma^{39},~~
 \gamma^{349},~~
 a\gamma^{569}+b\gamma^{478},
 a\gamma^{789}+b\gamma^{456},~~
 a\gamma^{1234}-b\gamma^{4},
\nonumber\\&&
 a\gamma^{3456}-b\gamma^{3789},~~
 \gamma^{5678},~~
 \gamma^{1256},~~
 \gamma^{1278},~~
 a\gamma^{3478}-b\gamma^{3569},
\label{matrices-3}
\end{eqnarray}
where $a=\cos^2\theta-\sin^2\theta$ and $b=2\cos\theta\sin\theta$.
We have used the fact $\gamma^{123456789}=1$. For $12\partial_-U_{(m)}
+V_{(m)}$ in $D_{(m)}$ be expanded in terms of (\ref{matrices-3}),
$\theta$, $U_{(m)}$ and $V_{(m)}$ must take the form
\begin{eqnarray}
\theta&=&
 a_1\gamma^{129}
 +a_2\gamma^{349}
 +a_3\gamma^{569}
 +a_4\gamma^{478}
 +a_5\gamma^{789}
 +a_6\gamma^{456}
 +a_7\gamma^{124},\\
U_{(m)}&=&
\alpha^m_1\gamma^{129}
+\alpha^m_2\gamma^{349}
+\alpha^m_3\gamma^{569}
+\alpha^m_4\gamma^{478}
+\alpha^m_5\gamma^{789}
+\alpha^m_6\gamma^{456}
+\alpha^m_7\gamma^{124} ,\\
V_{(m)}&=&
\mu^m_1\gamma^{12}
+\mu^m_2\gamma^{34}
+\mu^m_3\gamma^{56}
+\mu^m_4\gamma^{78}
+\mu^m_5\gamma^{39},
\end{eqnarray}
where
\begin{eqnarray}
\alpha_1&=&(4a_1,4a_1,-2a_1,-2a_1,-2a_1,-2a_1,-2a_1,-2a_1,4a_1),\nn\\
\alpha_2&=&(-2a_2,-2a_2,4a_2,4a_2,-2a_2,-2a_2,-2a_2,-2a_2,4a_2),\nn\\
\alpha_3&=&(-2a_3,-2a_3,-2a_3,-2a_3,4a_3,4a_3,-2a_3,-2a_3,4a_3),\nn\\
\alpha_4&=&(-2a_4,-2a_4,-2a_4,4a_4,-2a_4,-2a_4,4a_4,4a_4,-2a_4),\nn\\
\alpha_5&=&(-2a_5,-2a_5,-2a_5,-2a_5,-2a_5,-2a_5,4a_5,4a_5,4a_5),\nn\\
\alpha_6&=&(-2a_6,-2a_6,-2a_6,4a_6,4a_6,4a_6,-2a_6,-2a_6,-2a_6),\nn\\
\alpha_7&=&(4a_7,4a_7,-2a_7,4a_7,-2a_7,-2a_7,-2a_7,-2a_7,-2a_7).
\end{eqnarray}
The condition that $D_{(m)}$ is expanded in terms of (\ref{matrices-3})
implies that
\begin{eqnarray}
U_{(m)}\theta+3\theta_{(m)}U_{(m)}=U_{(m)}^2+[U_{(m)},\theta]
\end{eqnarray}
must be expanded in terms of (\ref{matrices-3}).
Because
\begin{eqnarray}
U_{(m)}^2&=&
-(\alpha^m_1)^2-(\alpha^m_2)^2- \cdots -(\alpha^m_7)^2
+2\alpha^m_1\alpha_2^m\gamma^{1234}
+2\alpha^m_1\alpha^m_3\gamma^{1256}
+2\alpha^m_1\alpha^m_5\gamma^{1278}
\nonumber\\&&
+2\alpha^m_2\alpha^m_3\gamma^{3456}
-2\alpha^m_2\alpha^m_4\gamma^{3789}
+2\alpha^m_2\alpha^m_5\gamma^{3478}
-2\alpha^m_2\alpha^m_6\gamma^{3569}
-2\alpha^m_2\alpha^m_7\gamma^{1239}
\nonumber\\&&
+2\alpha^m_3\alpha^m_5\gamma^{5678}
+2\alpha^m_4\alpha^m_6\gamma^{5678}
+2\alpha^m_4\alpha^m_7\gamma^{1278}
+2\alpha^m_6\alpha^m_7\gamma^{1256},\\
{[}U_{(m)},\theta]&=&
2(\alpha^m_1a_4-a_1\alpha^m_4+\alpha^m_5a_7-a_5\alpha^m_7)\gamma^{356}
+2(\alpha^m_1a_6-a_1\alpha^m_6+\alpha^m_3a_7-a_3\alpha^m_7)\gamma^{378}
\nonumber\\&&
+2(\alpha^m_1a_7-a_1\alpha^m_7+\alpha^m_3a_4-a_3\alpha^m_4
 +\alpha^m_3a_6-a_3\alpha^m_6-\alpha^m_4a_5+a_4\alpha^m_5)\gamma^{49}
\nonumber\\&&
+2(\alpha^m_5a_6-a_5\alpha^m_6)\gamma^{123},
\end{eqnarray}
this condition means
\begin{eqnarray}
&&
\alpha^m_1\alpha^m_2=0,
\\
&&
\alpha^m_2\alpha^m_7=0,
\\&&
\alpha^m_1a_4-a_1\alpha^m_4+\alpha^m_5a_7-a_5\alpha^m_7=0,
\\&&
\alpha^m_1a_6-a_1\alpha^m_6+\alpha^m_3a_7-a_3\alpha^m_7=0,
\\&&
\alpha^m_1a_7-a_1\alpha^m_7+\alpha^m_3a_4-a_3\alpha^m_4
 +\alpha^m_3a_6-a_3\alpha^m_6-\alpha^m_4a_5+a_4\alpha^m_5=0,
\\&&
\alpha^m_5a_6-a_5\alpha^m_6=0,
\end{eqnarray}
which are solved by one of the followings
\begin{itemize}
  \item $a_1,~a_3,~a_5\neq 0$, others = 0
  \item $a_2,~a_3,~a_5\neq 0$, others = 0
  \item $a_2,~a_4,~a_6\neq 0$, others = 0
  \item $a_4,~a_6,~a_7\neq 0$, others = 0
\end{itemize}
These all cases are contained in the cases, $a=0$ or $b=0$ in
(\ref{matrices-3}), and thus reduce to the simplest case considered in the
previous subsection. It is interesting to examine whether
the condition that $D_{(m)}$ is expanded completely in terms
of projectors leads to the pp-wave background (\ref{pp})
for general Cartan matrices.

\sect{Super-isometry algebra}

In this section, we examine the super-isometry algebra in
the background~(\ref{pp}).

Killing vector fields of the metric are solutions of the Killing vector
equations, $\CL_\xi g_{\mu\nu}=0$.
For the metric~(\ref{pp}), the Killing vector equations are
\begin{eqnarray}
\partial_+\xi_+=0,&&
\partial_m\xi_n+\partial_n\xi_m=0,\nonumber\\
\partial_+\xi_-+\partial_-\xi_+=0,&&
\partial_+\xi_m+\partial_m\xi_+=0,\nonumber\\
\partial_-\xi_-+A_{m}x^m\xi_m=0,&&
\partial_-\xi_m+\partial_m\xi_--2A_{m}x^m\xi_+=0.
\end{eqnarray}
Solutions for these equations contain several integral constants,
which represent individual solutions.
Each solution corresponds to components of a Killing vector field.
One finds that the Killing vectors are
\begin{eqnarray}
&&
\xi_{e_+}=-\partial_+,~~~
\xi_{e_-}=-\partial_-,~~~
\xi_{M_{pq}}=x^p\partial_q-x^q\partial_p,
\nonumber\\&&
\xi_{e_m}=-\sqrt{-A_m}\sin(\sqrt{-A_m}x^-)x^m\partial_+
 -\cos(\sqrt{-A_m}x^-)\partial_m,
\nonumber\\&&
\xi_{e_m^*}= -A_m\cos(\sqrt{-A_m}x^-)x^m\partial_+
 -\sqrt{-A_m}\sin(\sqrt{-A_m}x^-)\partial_m,
 \label{KV}
\end{eqnarray}
where $\xi_{M_{pq}}$ exists only when $A_p=A_q$.\footnote{Note that $A_m<0$
from eqs.~\p{A_m} and \p{A2}.}
The isometry algebra is obtained from these expressions as
\begin{eqnarray}
&&\[e_-,e_m]=e_m^*,~~~
\[e_-,e_m^*]=A_me_m,~~~
\[e_m,e_n^*]=-A_m\delta_{mn}e_+,\nonumber\\
&&\[M_{pq},e_m]=2\delta_{qm}e_p,~~~
\[M_{pq},e_m^*]=2\delta_{qm}e_p^*,~~~~\mbox{iff}~~A_m=A_p=A_q,\nonumber\\
&&\[M_{mn},M_{pq}]=4\delta_{np}M_{mq},~~~~\mbox{iff}~~A_m=A_n=A_p=A_q.
\label{ee}
\end{eqnarray}
The first line shows the typical structure shared among isometries of
pp-wave backgrounds, nine-dimensional Heisenberg algebra generated by
$e_m$ and $e_m^*$ with an outer-automorphism $e_-$.
The Lorentz algebra is a direct sum of algebras which are generated by
individual sets of generators, $\{M_{ij}\}$, $\{M_{i'j'}\}$,...
with $A_i=A_j\neq A_{i'}=A_{j'}\neq ...$. The flux $F$ may break the
Lorentz symmetry further. Instead of examining Lie derivative of $F$ here,
we will examine this in the course of deriving the super-isometry algebra,
where the Lorentz generator $M_{pq}$ is restricted by the consistency
of the algebra.

Let us examine the super-isometry algebra of the background.
The extra Killing spinor is determined by Killing spinor equations as
a solution of $[U_{(m)}^2-12^2A_m]\varepsilon_-=0$.
The extra Killing spinors, $\PP_I\varepsilon_-$,
lie in the $2N$-dimensional subspace, on which $U_{(m)}$ is of the form
\begin{eqnarray}
U_{(m)}=12\sqrt{-A_m}J,~~~~~
J=i\PP,~~~~
\PP=\sum_{I=I_1,...,I_{2N}}\PP_I.
\end{eqnarray}
One finds that the Killing spinor takes the form
\begin{eqnarray}
&&\varepsilon=\left(
  \begin{array}{c}
    \chi_++\frac{\sqrt{2}}{2}\sqrt{-A_m}x^m\gamma_mJ\chi_-   \\
    \chi_-   \\
  \end{array}
\right),~~~
  \begin{array}{l}
\chi_+=e^{-\frac{\theta}{4}x^-}\psi_+,       \\
\chi_-=e^{\frac{\theta}{12}x^-}\psi_-,      \\
  \end{array}
~~~\PP\psi_-=\psi_-.
\end{eqnarray}

First, we examine the commutation relations between two of supercharges.
To do this, we calculate $\bar\varepsilon_1(\psi_+,\psi_-)$
$\Gamma^{\hat \mu}\varepsilon_2(\psi_+',\psi_-')\partial_{\hat\mu}$,
where $\bar\varepsilon=\varepsilon^TC$ and the index with a hat represents
the curved index. In accordance with~(\ref{gamma}),
the even-dimensional charge conjugation matrix $C$ is expressed as
$C=c\otimes i\sigma_2$ where $c$ is the nine-dimensional charge conjugation
matrix. One finds that the result is rewritten compactly in terms of Killing
vectors~(\ref{KV}):
\begin{eqnarray}
\bar\varepsilon_1\Gamma^{\hat\mu}\varepsilon_2\partial_{\hat\mu}&=&
-\sqrt{2}\bar\psi_+\psi_+'\xi_{e_+}
+\sqrt{2}\bar\psi_-\PP\psi_-'
\xi_{e_-}
+\frac{\sqrt{2}}{48}\bar\psi_-\PP\{\gamma^{mn},U_{(n)}\}\PP
\psi_-'\xi_{M_{mn}}
\\
&&+[~\bar\psi_+\gamma^m\PP\psi_-'+\bar\psi_-\PP\gamma^m\psi_+']
\xi_{e_m}
+[\bar\psi_+\gamma^mJ\psi_-'-\bar\psi_-J\gamma^m\psi_+']
 \frac{1}{\sqrt{-A_m}}\xi_{e^*_m}.\nonumber
\end{eqnarray}
The QQ anti-commutators can be read off as
\begin{eqnarray}
&&
\{Q_+,Q_+\}=-\sqrt{2}c~e_+,~~~
\{Q_-,Q_-\}=\sqrt{2}c\PP ~e_-
+\frac{\sqrt{2}}{48}c\PP\{\gamma^{mn},U_{(n)}\}\PP
~M_{mn},\label{QQ}\\
&&\{Q_+,Q_{-}\}=
c\gamma^m\PP~e_m
+c\gamma^mJ\frac{1}{\sqrt{-A_m}}~e_m^*,~~~
\{Q_-,Q_{+}\}=
c\gamma^m\PP~e_m
-cJ\gamma^m\frac{1}{\sqrt{-A_m}}~e_m^*.
\nonumber
\end{eqnarray}
The second term in the right hand side of the second equation survives
only when $\PP\{\gamma^{mn},U_{(n)}\}\PP\neq 0$. This enables us to know
the actual unbroken Lorentz symmetry in the presence of the flux.

Secondly, we examine the commutation relations between bosonic generators
and a supercharge. For this purpose, we define the spinorial Lie derivative
$\LL_\xi$ \cite{Townsend:Killing spinors, F:On the supersymmetries,
Ortin:A note} along a Killing vector field $\xi$
\begin{eqnarray}
\LL_\xi
=\xi^{\hat\mu}\nabla_{\hat\mu}
+\frac{1}{4}\nabla_{\hat\mu}\xi_{\hat\nu}\Gamma^{\hat\mu\hat\nu}
\end{eqnarray}
where the covariant derivative $\nabla_{\hat\mu}$ acting on spinors (vectors)
contains the spin (Levi-Civita) connection. After some algebra, we find that
\begin{eqnarray}
&&
\LL_{\xi_{e_+}}\varepsilon(\psi_+,\psi_-)=0,~~~
\LL_{\xi_{e_-}}\varepsilon(\psi_+,\psi_-)=
\varepsilon(\frac{\theta}{4}\psi_+,-\frac{\theta}{12}\psi_-),
\nonumber\\&&
\LL_{\xi_{e_m}}\varepsilon(\psi_+,\psi_-)=
\varepsilon(-\frac{\sqrt{2}}{24}\gamma^mU_{(m)}\psi_-, 0),~~~
\LL_{\xi_{e_m^*}}\varepsilon(\psi_+,\psi_-)=
\varepsilon(-\frac{\sqrt{2}}{2}\gamma^mA_m\psi_-, 0),
\nonumber\\&&
\LL_{\xi_{M_{pq}}}\varepsilon(\psi_+,\psi_-)=
\varepsilon(\frac{1}{2}\gamma^{pq}\psi_+,\frac{1}{2}\gamma^{pq}\psi_-).
\label{KS}
\end{eqnarray}
The last equation is satisfied only when $U_{(p)}=U_{(q)}$,
which is always satisfied on the $2N$-dimensional subspace.
The commutation relations can be read off from~(\ref{KS}) as
\begin{eqnarray}
&&
\[e_+,Q_{\pm}]=0,~~~
\[e_-,Q_+]=\frac{1}{4}Q_+{\theta},~~~
\[e_-,Q_-]=-\frac{1}{12}Q_-{\theta},
\nonumber\\&&
\[e_m,Q_-]=-\frac{\sqrt{2}}{2}\sqrt{-A_m}Q_+\gamma_mJ,~~~
\[e_m,Q_+]=0,
\label{eQ}\\&&
\[e_m^*,Q_-]=-\frac{\sqrt{2}}{2}A_mQ_+\gamma_m,~~~
\[e_m^*,Q_+]=0,~~~
\[M_{pq},Q_\pm]=\frac{1}{2}Q_\pm\gamma_{pq}.\nonumber
\end{eqnarray}

In summary, the super-isometry algebra of~(\ref{pp}) has been obtained
as (\ref{ee}), (\ref{eQ}) and (\ref{eQ}). The super-isometry algebras
have been given in \cite{Figueroa-O'Farrill:Homogeneous}
for 32 supersymmetric case and in \cite{Michelson:A pp-wave} for 26
supersymmetric case. Our superalgebra covers all supersymmetric cases.
Note that the superalgebra obtained above preserves $\Omega$-charge
\begin{eqnarray}
  \begin{array}{c|c|c|c|c|c|c}
 e_m &e_+&e_- &e_m^* &M_{mn}&Q_+&Q_-
 \\\hline
    1& 2 &0   &1     &0     &1  &0
  \end{array},
\end{eqnarray}
and thus
enjoys $\Omega$-grading property
which has played a crucial role in the Penrose limit~\cite{HKS:Super-pp-wave}.


\sect{Summary and Discussions}

We have established a uniqueness theorem which states that
any M-theory pp-wave background can be reduced to the form~(\ref{pp-}) modulo
coordinate transformations, if there exists at least one non-harmonic
extra Killing spinor. We have examined further the cases in which $D_{(m)}$
is expanded in terms of projectors only. For the cases in which projectors are
characterized by monomial Cartan matrices, (\ref{matrices-1}) and
(\ref{matrices-2}), we have found that the background reduces to the form
(\ref{pp}). We have also discussed more general cases in which Cartan
matrices are not monomials, and found that the background reduces to the
form (\ref{pp}) again. This observation suggests that the background
reduces to the form (\ref{pp}) if the Killing spinors are characterized
by projectors only. In fact, we have showed in appendix B that
the Killing spinors on the time-dependent pp-wave background related to
the anti-Mach type background~\cite{BMO} are not characterized by
projectors only. It is expected that our observation may be useful in
constructing time-dependent backgrounds.

In addition, for the pp-wave background~(\ref{pp}), we have derived the
super-isometry algebras which contain 18, 20, 22, 24, 26 and 32 supercharges.

It is interesting to examine the similar uniqueness theorem
for pp-wave backgrounds in lower dimensions.
For type-IIB pp-wave backgrounds with the self-dual five-form RR-field
strength, the similar uniqueness theorem can be discussed \cite{S:IIB}.
We also expect that the similar uniqueness theorems can be established
for pp-wave backgrounds in six-, five- and four-dimensions.

It is known that Killing spinor equations for maximally supersymmetric
backgrounds imply the supergravity equations of motion. We have seen that
the pp-wave backgrounds which admit at least one extra Killing spinor
automatically satisfy the supergravity equation of motion,
and thus the Killing spinor equations imply the supergravity equation
of motion even for non-maximally supersymmetric cases.
This suggests that Killing spinor equations for backgrounds with
extra supersymmetries have rich algebraic structures,
just as those for maximally supersymmetric backgrounds.
In \cite{FP;Maximally},
maximally supersymmetric backgrounds were classified
examining the algebraic structures of Killing spinors.\footnote{
A classification of backgrounds which admit at least one Killing spinor
was discussed in \cite{GP;Geometry of D=11}
}
It is interesting to classify all non-maximally supersymmetric backgrounds
which admit more than 16 supersymmetries.

M-brane actions on the non-maximally supersymmetric pp-wave backgrounds
can be constructed using supercurrents on the supergroup manifold
corresponding to the superalgebras obtained in section 4.
Examining the properties of such models may be useful to gain deeper
insights to M-theory and the non-maximally supersymmetric pp-wave backgrounds.

\section*{Acknowledgements}
The work of NO was supported in part by Grants-in-Aid for Scientific Research
Nos. 12640270 and 02041.

\appendix

\bigskip
\bigskip

\top
\textbf{\Large Appendix}
\section{the case (\ref{matrices-2})}

In this appendix, we discuss the case in which $D_{(m)}$
is expanded with respect to Cartan matrices (\ref{matrices-2}).
The $\theta=\frac{1}{3!}\xi_{lmn}\gamma^{lmn}$ takes the form
\begin{eqnarray}
\theta=
 b_1\gamma^{123}
 +b_2\gamma^{145}
 +b_3\gamma^{167}
 +b_4\gamma^{246}
 +b_5\gamma^{257}
 +b_6\gamma^{347}
 +b_7\gamma^{356},
 \label{theta2:a}
\end{eqnarray}
and
$A_{mn}$ is restricted to be non-vanishing only when
$m=n$ and $(m,n)=(8,9)$.
Because terms in (\ref{theta2:a}) commute with each other,
(\ref{KSE:before}) reduces to (\ref{KSE}) as before.
$U_{(m)}$ and $V_{(m)}$ for~\p{matrices-2} is expressed as
\begin{eqnarray}
U_{(m)}
&=&\beta^m_1\gamma^{123}
+\beta^m_2\gamma^{145}
+\beta^m_3\gamma^{167}
+\beta^m_4\gamma^{246}
+\beta^m_5\gamma^{257}
+\beta^m_6\gamma^{347}
+\beta^m_7\gamma^{356}
\nn\\
V_{(m)}&=&
 \nu^m\gamma^{89}\label{U2}
\end{eqnarray}
where
\begin{eqnarray}
\beta_1&=&(4b_1,4b_1,4b_1,-2b_1,-2b_1,-2b_1,-2b_1,-2b_1,-2b_1),
\nonumber\\
\beta_2&=&(4b_2, -2b_2, -2b_2, 4b_2, 4b_2, -2b_2, -2b_2, -2b_2, -2b_2),\nn\\
\beta_3&=&(4b_3, -2b_3, -2b_3, -2b_3, -2b_3, 4b_3, 4b_3, -2b_3, -2b_3),\nn\\
\beta_4&=&(-2b_4, 4b_4, -2b_4, 4b_4, -2b_4, 4b_4, -2b_4, -2b_4, -2b_4),\nn\\
\beta_5&=&(-2b_5, 4b_5, -2b_5, -2b_5, 4b_5, -2b_5, 4b_5, -2b_5, -2b_5),\nn\\
\beta_6&=&(-2b_6, -2b_6, 4b_6, 4b_6, -2b_6, -2b_6, 4b_6, -2b_6, -2b_6),\nn\\
\beta_7&=&(-2b_7, -2b_7, 4b_7, -2b_7, 4b_7, 4b_7, -2b_7, -2b_7, -2b_7),
\nn\\
\nu^8&=&-\nu^9=-12^2A_{89}.\label{beta}
\end{eqnarray}

In this case, all the matrices (\ref{matrices-2})
can be expressed as products of four matrices, $\gamma^{89}$,
$\gamma^{123}$, $\gamma^{167}$ and $\gamma^{356}$. From these matrices,
we make four rank-8 projection operators
\begin{eqnarray}
&&
P_0=\frac{1}{2}(\1+i\gamma^{89}),~~~
P_1=\frac{1}{2}(\1+i\gamma^{123}),\nonumber\\&&
P_2=\frac{1}{2}(\1+i\gamma^{167}),~~~
P_3=\frac{1}{2}(\1+i\gamma^{356}),
\label{P2}
\end{eqnarray}
which satisfy
\begin{eqnarray}
P_A^2=P_A,~~~P_AP_B=P_BP_A,~~~
A,B=0,1,2,3.
\end{eqnarray}
Because
\begin{eqnarray}
&&
\gamma^{123}=-i(2P_1-\1),~~~
\gamma^{145}=-i(2P_0-\1)(2P_1-\1)(2P_2-\1),~~~
\gamma^{167}=-i(2P_2-\1),
\nonumber\\&&
\gamma^{246}=i(2P_0-\1)(2P_2-\1)(2P_3-\1),~~~
\gamma^{257}=-i(2P_1-\1)(2P_2-\1)(2P_3-\1),
\\&&
\gamma^{347}=-i(2P_0-\1)(2P_1-\1)(2P_3-\1),~~~
\gamma^{356}=-i(2P_3-\1),~~~
\gamma^{89}=-i(2P_0-1), \nonumber
\end{eqnarray}
$U_{(m)}$ and $V_{(m)}$ in (\ref{U2})
can be expressed in terms of projection operators $P_A$
as
\begin{eqnarray}
U_{(m)}&=&
i(\beta^m_1+\beta^m_2+\beta^m_3-\beta^m_4+\beta^m_5+\beta^m_6+\beta^m_7)\1
\nonumber\\&&
-2i(\beta^m_2-\beta^m_4+\beta^m_6)P_0
-2i(\beta^m_1+\beta^m_2+\beta^m_5+\beta^m_6)P_1
\nonumber\\&&
-2i(\beta^m_2+\beta^m_3-\beta^m_4+\beta^m_5)P_2
-2i(-\beta^m_4+\beta^m_5+\beta^m_6+\beta^m_7)P_3
\nonumber\\&&
+4i(\beta^m_2+\beta^m_6)P_0P_1
+4i(\beta^m_2-\beta^m_4)P_0P_2
+4i(-\beta^m_4+\beta^m_6)P_0P_3
\nonumber\\&&
+4i(\beta^m_2+\beta^m_5)P_1P_2
+4i(\beta^m_5+\beta^m_6)P_1P_3
+4i(-\beta^m_4+\beta^m_5)P_2P_3
\nonumber\\&&
-8i\beta^m_2P_0P_1P_2
-8i\beta^m_6P_0P_1P_3
+8i\beta^m_4P_0P_2P_3
-8i\beta^m_5P_1P_2P_3,
\nonumber\\
V_{(m)}&=&
i\nu^m-2i\nu^mP_0
\end{eqnarray}
Introduce rank-1 projectors of a 16-component spinor into the $I$-th component:
\begin{eqnarray}
\PP_I=diag(0,...,0,\stackrel{I}{1},0,...,0),
\end{eqnarray}
and the rank-8 projection operators $P_A$ can be obtained by their linear
combinations as
\begin{eqnarray}
&&
P_0=\sum_{I=1,2,3,4,5,6,7,8}\PP_I,~~~
P_1=\sum_{I=1,2,3,4,9,10,11,12}\PP_I,
\nonumber\\&&P_2=\sum_{I=1,2,5,6,9,10,13,14}\PP_I,~~~
P_3=\sum_{I=1,3,5,7,9,11,13,15}\PP_I.
\label{PP2}
\end{eqnarray}
In terms of $\PP_I$, (\ref{KSE}) becomes
\begin{eqnarray}
&\Bigl[&
\Big(
12i\partial_-(\beta_1^m+\beta_2^m+\beta_3^m-\beta_4^m+\beta_5^m+\beta_6^m
+\beta_7^m)+i\nu^m
\nonumber\\&&~~~
-(\beta_1^m+\beta_2^m+\beta_3^m-\beta_4^m+\beta_5^m+\beta_6^m+\beta_7^m)^2
-12^2A_m\Big)~\1
\nonumber\\&&
+\Big(-24i\partial_-(
\beta_1^m+\beta_2^m+\beta_3^m-\beta_4^m+\beta_5^m+\beta_6^m+\beta_7^m
)
-2i\nu^m
\Big)~\PP_1\nonumber\\&&
+\Big(
-24i\partial_-(\beta_1^m+\beta_2^m+\beta_3^m)
-2i\nu^m
+4(\beta^m_1+\beta^m_2+\beta^m_3)(-\beta^m_4+\beta^m_5+\beta^m_6+\beta^m_7)
\Big)~\PP_2\nonumber\\&&
+\Big(
-24i\partial_-(\beta^m_1+\beta^m_6+\beta^m_7
)
-2i\nu^m
+4(\beta^m_1+\beta^m_6+\beta^m_7)(\beta^m_2+\beta^m_3-\beta^m_4+\beta^m_5)
\Big)~\PP_3\nonumber\\&&
+\Big(
-24i\partial_-(\beta^m_1-\beta^m_4+\beta^m_5
)
-2i\nu^m
+4(\beta^m_1-\beta^m_4+\beta^m_5)(\beta^m_2+\beta^m_3+\beta^m_6+\beta^m_7)
\Big)~\PP_4\nonumber\\&&
+\Big(
-24i\partial_-(\beta^m_3-\beta^m_4+\beta^m_7
)
-2i\nu^m
+4(\beta^m_3-\beta^m_4+\beta^m_7)(\beta^m_1+\beta^m_2+\beta^m_5+\beta^m_6)
\Big)~\PP_5\nonumber\\&&
+\Big(
-24i\partial(\beta^m_3+\beta^m_5+\beta^m_6
)
-2i\nu^m
+4(\beta^m_3+\beta^m_5+\beta^m_6)(\beta^m_1+\beta^m_2-\beta^m_4+\beta^m_7)
\Big)~\PP_6\nonumber\\&&
+\Big(
-24i\partial_-(\beta^m_2+\beta^m_5+\beta^m_7
)
-2i\nu^m
+4(\beta^m_2+\beta^m_5+\beta^m_7)(\beta^m_1+\beta^m_3-\beta^m_4+\beta^m_6)
\Big)~\PP_7\nonumber\\&&
+\Big(
-24i\partial_-(\beta^m_2-\beta^m_4+\beta^m_6
)
-2i\nu^m
+4(\beta^m_2-\beta^m_4+\beta^m_6)(\beta^m_1+\beta^m_3+\beta^m_5+\beta^m_7)
\Big)~\PP_8\nonumber\\&&
+\Big(
-24i\partial_-(\beta^m_1+\beta^m_3+\beta^m_5+\beta^m_7
)
+4(\beta^m_2-\beta^m_4+\beta^m_6)(\beta^m_1+\beta^m_3+\beta^m_5+\beta^m_7)
\Big)~\PP_9\nonumber\\&&
+\Big(
-24i\partial_-(\beta^m_1+\beta^m_3-\beta^m_4+\beta^m_6
)
+4(\beta^m_2+\beta^m_5+\beta^m_7)(\beta^m_1+\beta^m_3-\beta^m_4+\beta^m_6)
\Big)~\PP_{10}\nonumber\\&&
+\Big(
-24i\partial_-(\beta^m_1+\beta^m_2-\beta^m_4+\beta^m_7
)
+4(\beta^m_3+\beta^m_5+\beta^m_6)(\beta^m_1+\beta^m_2-\beta^m_4+\beta^m_7)
\Big)~\PP_{11}\nonumber\\&&
+\Big(
-24i\partial_-(\beta^m_1+\beta^m_2+\beta^m_5+\beta^m_6
)
+4(\beta^m_3-\beta^m_4+\beta^m_7)(\beta^m_1+\beta^m_2+\beta^m_5+\beta^m_6)
\Big)~\PP_{12}\nonumber\\&&
+\Big(
-24i\partial_-(\beta^m_2+\beta^m_3+\beta^m_6+\beta^m_7
)
+4(\beta^m_1-\beta^m_4+\beta^m_5)(\beta^m_2+\beta^m_3+\beta^m_6+\beta^m_7)
\Big)~\PP_{13}\nonumber\\&&
+\Big(
-24i\partial_-(\beta^m_2+\beta^m_3-\beta^m_4+\beta^m_5
)
+4(\beta^m_1+\beta^m_6+\beta^m_7)(\beta^m_2+\beta^m_3-\beta^m_4+\beta^m_5)
\Big)~\PP_{14}\nonumber\\&&
+\Big(
-24i\partial_-(-\beta^m_4+\beta^m_5+\beta^m_6+\beta^m_7
)
\nonumber\\&&\hspace{30mm}
+4(\beta^m_1+\beta^m_2+\beta^m_3)(-\beta^m_4+\beta^m_5+\beta^m_6+\beta^m_7)
\Big)~\PP_{15}
~~\Bigr]\varepsilon_-=0.
\end{eqnarray}
Again the coefficient of $\1$ of the above equation must vanish in order to
give an extra Killing spinor, which is then $\PP_{16}\varepsilon_-$. We get
\begin{eqnarray}
&&
12\partial_-(\beta_1^m+\beta_2^m+\beta_3^m-\beta_4^m+\beta_5^m+\beta_6^m
+\beta_7^m) +\nu^m=0,
\\&&
A_m=-\frac{1}{12^2}
(\beta_1^m+\beta_2^m+\beta_3^m-\beta_4^m+\beta_5^m+\beta_6^m+\beta_7^m)^2.
\label{A2}
\end{eqnarray}
The former equation implies that $\nu^m=0$, which means $A_{89}=0$,
and
\begin{eqnarray}
\partial_-(\beta_1^m+\beta_2^m+\beta_3^m-\beta_4^m+\beta_5^m+\beta_6^m
+\beta_7^m)=0,
\label{a2}
\end{eqnarray}
because $\beta_i^8=\beta_i^9$ while $\nu^8=-\nu^9$.
We see from eq.~\p{a2} that $A_m$ in \p{A2} is independent of $x^-$.
With the help of (\ref{beta}), eq.~\p{a2} leads to
\begin{eqnarray}
\partial_-b_1=\partial_-b_2=\partial_-b_3=\partial_-b_4=\partial_-b_5
=\partial_-b_6
=\partial_-b_7=0,
\end{eqnarray}
which implies that $\xi$ is independent of $x^-$.
The extra Killing spinors are determined as non-trivial solutions of
\begin{eqnarray}
&&\Bigl[
\Big(
-(\beta_1^m+\beta_2^m+\beta_3^m-\beta_4^m+\beta_5^m+\beta_6^m+\beta_7^m)^2
-12^2A_m\Big)~\1
\nonumber\\&&~~
+4(\beta^m_1+\beta^m_2+\beta^m_3)(-\beta^m_4+\beta^m_5+\beta^m_6+\beta^m_7)
~(\PP_2+\PP_{15}) \nonumber\\&&~~
+4(\beta^m_1+\beta^m_6+\beta^m_7)(\beta^m_2+\beta^m_3-\beta^m_4+\beta^m_5)
~(\PP_3+\PP_{14}) \nonumber\\&&~~
+4(\beta^m_1-\beta^m_4+\beta^m_5)(\beta^m_2+\beta^m_3+\beta^m_6+\beta^m_7)
~(\PP_4+\PP_{13}) \nonumber\\&&~~
+4(\beta^m_3-\beta^m_4+\beta^m_7)(\beta^m_1+\beta^m_2+\beta^m_5+\beta^m_6)
~(\PP_5+\PP_{12})
\nonumber\\&&~~
+4(\beta^m_3+\beta^m_5+\beta^m_6)(\beta^m_1+\beta^m_2-\beta^m_4+\beta^m_7)
~(\PP_6+\PP_{11})\nonumber\\&&~~
+4(\beta^m_2+\beta^m_5+\beta^m_7)(\beta^m_1+\beta^m_3-\beta^m_4+\beta^m_6)
~(\PP_7+\PP_{10})\nonumber\\&&~~
+4(\beta^m_2-\beta^m_4+\beta^m_6)(\beta^m_1+\beta^m_3+\beta^m_5+\beta^m_7)
~(\PP_8+\PP_9)
~~\Bigr]\varepsilon_-=0.
\end{eqnarray}
This again shows the two-fold degeneracy of the extra Killing spinors.
If (\ref{A2}) is satisfied, $\PP_1\varepsilon_-$ and $\PP_{16}\varepsilon_-$
become a pair of extra Killing spinors, and the background admits 18
Killing spinors, 16 standard and 2 extra Killing spinors.
In addition, if the coefficient of $(\PP_I+\PP_{17-I})$
vanishes, $\PP_I\varepsilon_-$ and $\PP_{17-I}\varepsilon_-$
become a pair of additional Killing spinors.
Examining these conditions, one finds
pp-wave backgrounds which admit 18, 20, 22, 24, 26, 32
Killing spinors \cite{GH:pp-waves in 11-dimensions,Michelson:A pp-wave}.

The supergravity equation of motion is automatically satisfied for pp-wave
backgrounds with extra supersymmetries, because
one sees from (\ref{A2}) and (\ref{beta}) that
\begin{eqnarray}
\triangle H=\sum_{m=1}^92A_m=-b_1^2-b_2^2-b_3^2-b_4^2-b_5^2-b_6^2-b_7^2,
\end{eqnarray}
and from (\ref{theta2:a}) that
\begin{eqnarray}
-\frac{1}{3!}\xi_{lmn}\xi^{lmn}=-b_1^2-b_2^2-b_3^2-b_4^2-b_5^2-b_6^2-b_7^2.
\end{eqnarray}

\bigskip

In summary, we have shown in this section that M-theory pp-wave backgrounds
which admit extra Killing spinors characterized by Cartan matrices
(\ref{matrices-2}) can be reduced again to the form~\p{pp}.

\section{Similarity transformations}
\bigskip

We study the similarity transformations of (\ref{matrices-1}) by
\begin{eqnarray}
S=e^{\theta\gamma^{178}}
~~~\mbox{ and then }~~~S'=e^{\theta'\gamma^{12}}.
\end{eqnarray}
The matrices (\ref{matrices-1}) become
\begin{eqnarray}
&&
a\gamma^{9}+b(a'\gamma^{1789}-b'\gamma^{2789}),~~~
a\gamma^{12}+b(a'\gamma^{278}+b'\gamma^{178}),~~~
\gamma^{34},~~~
\gamma^{56},~~~
\gamma^{78},~~~
\gamma^{129},
\nonumber\\&&
a\gamma^{349}+b(a'\gamma^{256}+b'\gamma^{156}),~~~
a\gamma^{569}+b(a'\gamma^{234}+b'\gamma^{134}),~~~
a\gamma^{789}-b(a'\gamma^{19}-b'\gamma^{29}),
\nonumber\\&&
a\gamma^{1234}-b(a'\gamma^{1569}-b'\gamma^{2569}),~~~
a\gamma^{3456}-b(a'\gamma^{2}+b'\gamma^{1}),~~~
\gamma^{5678},
\nonumber\\&&
a\gamma^{1256}-b(a'\gamma^{1349}-b'\gamma^{2349}),~~~
a\gamma^{1278}-b(a'\gamma^{2}+b'\gamma^{1}),~~~
\gamma^{3478},
\label{matrices-4}
\end{eqnarray}
where $a=\cos^2\theta-\sin^2\theta$, $b=2\cos\theta\sin\theta$,
$a'=\cos^2\theta'-\sin^2\theta'$ and $b'=2\cos\theta'\sin\theta'$.
The Cartan matrices $SH_IS^{-1}$ contain the matrices used for
the anti-Mach type pp-wave\footnote{
The matrix theory on this background was constructed in
\cite{SY}
}
found in \cite{BMO},
and thus the matrices
(\ref{matrices-4}) is relevant to the time-dependent pp-wave\footnote{
Strictly speaking this background is not time-dependent
because there is a static chart.
In the Brinkmann coordinate system we are using,
the background looks time-dependent, but
the anti-Mach type background is static.
}
which is related to the anti-Mach type pp-wave
by a time-dependent coordinate transformation~\cite{HPW}.
It is interesting to examine the $D_{(m)}$ for
this time-dependent pp-wave. For this purpose, we consider the case in which
$\theta$ is expanded with respect to a part of (\ref{matrices-4}),
\begin{eqnarray}
\theta&=&
a_1\gamma^{278}
+a_2\gamma^{178}
+a_3\gamma^{129}
+a_4\gamma^{349}
+a_5\gamma^{256}
+a_6\gamma^{156},
\\
U_{(m)}&=&
\alpha^m_1\gamma^{278}
+\alpha^m_2\gamma^{178}
+\alpha^m_3\gamma^{129}
+\alpha^m_4\gamma^{349}
+\alpha^m_5\gamma^{256}
+\alpha^m\gamma^{156},
\\
V_{(m)}&=&
\mu^m\gamma^{12},
\end{eqnarray}
where
\begin{eqnarray}
\alpha_1&=&(-2a_1,4a_1,-2a_1,-2a_1,-2a_1,-2a_1,4a_1,4a_1,-2a_1),\nonumber\\
\alpha_2&=&(4a_2,-2a_2,-2a_2,-2a_2,-2a_2,-2a_2,4a_2,4a_2,-2a_2),\nonumber\\
\alpha_3&=&(4a_3,4a_3,-2a_3,-2a_3,-2a_3,-2a_3,-2a_3,-2a_3,4a_3),\nonumber\\
\alpha_4&=&(-2a_4,-2a_4,4a_4,4a_4,-2a_4,-2a_4,-2a_4,-2a_4,4a_4),\nonumber\\
\alpha_5&=&(-2a_5,4a_5,-2a_5,-2a_5,4a_5,4a_5,-2a_5,-2a_5,-2a_5),\nonumber\\
\alpha_6&=&(4a_6,-2a_6,-2a_6,-2a_6,4a_6,4a_6,-2a_6,-2a_6,-2a_6),\nonumber\\
\mu^1&=&-\mu^2=-12A_{12}.
\end{eqnarray}
It follows that
\begin{eqnarray}
U_{(m)}^2&=&
-2\alpha^m_1\alpha^m_3\gamma^{1789}
+2(\alpha^m_1\alpha^m_5+\alpha^m_2\alpha^m_6)\gamma^{5678}
+2\alpha^m_2\alpha^m_3\gamma^{2789}
+2\alpha^m_3\alpha^m_4\gamma^{1234}
\nonumber\\&&
-2\alpha^m_3\alpha^m_5\gamma^{1569}
+2\alpha^m_3\alpha^m_6\gamma^{2569},
\\
{[}U_{(m)},\theta]&=&
2(\alpha^m_1a_2-\alpha^m_2a_1+\alpha^m_5a_6-\alpha^m_6a_5)\gamma^{12}
-2(\alpha^m_1a_4-\alpha^m_4a_1)\gamma^{156}
\nonumber\\&&
+2(\alpha^m_1a_6-\alpha^m_6a_1+\alpha^m_2a_5-\alpha^m_5a_2)\gamma^{349}
+2(\alpha^m_2a_4-\alpha^m_4a_2)\gamma^{256}
\nonumber\\&&
+2(\alpha^m_4a_5-\alpha^m_5a_4)\gamma^{178}
-2(\alpha^m_4a_6-\alpha^m_6a_4)\gamma^{278},
\end{eqnarray}
and thus the $D_{(m)}$ becomes
\begin{eqnarray}
D_{(m)}&=&
(\mu^m +2(\alpha^m_1a_2-\alpha^m_2a_1+\alpha^m_5a_6-\alpha^m_6a_5)
)\gamma^{12}
\nonumber\\&&
+(\partial_-\alpha^m_1
-2(\alpha^m_4a_6-\alpha^m_6a_4)
)\gamma^{278}
+(\partial_-\alpha^m_2
+2(\alpha^m_4a_5-\alpha^m_5a_4) )\gamma^{178}
+\partial_-\alpha^m_3\gamma^{129}
\nonumber\\&&
+(\partial_-\alpha^m_4
 +2(\alpha^m_1a_6-\alpha^m_6a_1+\alpha^m_2a_5-\alpha^m_5a_2))\gamma^{349}
 \nonumber\\&&
+(\partial_-\alpha^m_5
 +2(\alpha^m_2a_4-\alpha^m_4a_2))\gamma^{256}
+(\partial_-\alpha^m
 -2(\alpha^m_1a_4-\alpha^m_4a_1))\gamma^{156}
 \nonumber\\&&
-2\alpha^m_1\alpha^m_3\gamma^{1789}
+2\alpha^m_2\alpha^m_3\gamma^{2789}
+2\alpha^m_3\alpha^m_4\gamma^{1234}
-2\alpha^m_3\alpha^m_5\gamma^{1569}
+2\alpha^m_3\alpha^m_6\gamma^{2569}
 \nonumber\\&&
+2(\alpha^m_1\alpha^m_5+\alpha^m_2\alpha^m_6)\gamma^{5678}.
\end{eqnarray}
Note that there is the term $-2\alpha^m_1\alpha^m_3\gamma^{1789}
+2\alpha^m_2\alpha^m_3\gamma^{2789}$, but not a $\gamma^9$ term.
If there was a $\gamma^9$ term, $D_{(m)}$ could be a Cartan. We find that
$D_{(m)}$ for this background is not expanded in terms of projectors only.



\begin{thebibliography}{99}
\bibitem{Metsaev:Type IIB}
R.~R.~Metsaev,
``Type IIB Green-Schwarz superstring in plane wave Ramond-Ramond  background,''
Nucl.\ Phys.\ B {\bf 625} (2002) 70
[arXiv:hep-th/0112044].
\bibitem{BFHP:A new maximally}
M.~Blau, J.~Figueroa-O'Farrill, C.~Hull and G.~Papadopoulos,
``A new maximally supersymmetric background of IIB superstring theory,''
JHEP {\bf 0201} (2002) 047
[arXiv:hep-th/0110242].
\bibitem{BMN}
D.~Berenstein, J.~M.~Maldacena and H.~Nastase,
``Strings in flat space and pp waves from N = 4 super Yang Mills,''
JHEP {\bf 0204} (2002) 013
[arXiv:hep-th/0202021],
and references thereof.
\bibitem{Penrose}
R.~Penrose,
``Any space-time has a plane wave as a limit'',
in Differential Geometry and Relativity, Cahen and Flato eds. (1976)
Reidel Publishing, Dordrecht-Holland.\\
R.~Gueven,
``Plane wave limits and T-duality,''
Phys.\ Lett.\ B {\bf 482} (2000) 255
[arXiv:hep-th/0005061].
\bibitem{Hull:Exact pp wave}
C.~M.~Hull,
``Exact pp wave solutions of 11-dimensional supergravity,''
Phys.\ Lett.\ B {\bf 139}, 39 (1984).
\bibitem{Kowalski-Glikman:vacuum}
J.~Kowalski-Glikman,
``Vacuum states in supersymmetric Kaluza-Klein theory,''
Phys.\ Lett.\ B {\bf 134}, 194 (1984).
\bibitem{Chrusciel:The isometry}
P.~T.~Chrusciel and J.~Kowalski-Glikman,
``The isometry group and Killing spinors for the pp wave space-time in
$D = 11$ supergravity,''
Phys.\ Lett.\ B {\bf 149}, 107 (1984).
\bibitem{Figueroa-O'Farrill:Homogeneous}
J.~Figueroa-O'Farrill and G.~Papadopoulos,
``Homogeneous fluxes, branes and a maximally supersymmetric solution of
M-theory,''
JHEP {\bf 0108}, 036 (2001)
[arXiv:hep-th/0105308].
\bibitem{BFCP:Penrose limits}
M.~Blau, J.~Figueroa-O'Farrill, C.~Hull and G.~Papadopoulos,
`Penrose limits and maximal supersymmetry,''
Class.\ Quant.\ Grav.\  {\bf 19} (2002) L87
[arXiv:hep-th/0201081].
\bibitem{HKS:Super-pp-wave}
M.~Hatsuda, K.~Kamimura and M.~Sakaguchi,
``Super-pp-wave algebra from super-AdS $\times$ S algebras in
eleven-dimensions,''
Nucl.\ Phys.\ B {\bf 637} (2002) 168
[arXiv:hep-th/0204002].
\bibitem{CLP:Penrose}
M.~Cvetic, H.~Lu and C.~N.~Pope,
``Penrose limits, pp-waves and deformed M2-branes,''
arXiv:hep-th/0203082.
\bibitem{Michelson:Twisted}
J.~Michelson,
``(Twisted) toroidal compactification of pp-waves,''
Phys.\ Rev.\ D {\bf 66} (2002) 066002
[arXiv:hep-th/0203140].
\bibitem{CLP:M-theory pp-waves}
M.~Cvetic, H.~Lu and C.~N.~Pope,
``M-theory pp-waves, Penrose limits and supernumerary supersymmetries,''
Nucl.\ Phys.\ B {\bf 644}, 65 (2002)
[arXiv:hep-th/0203229].
\bibitem{GH:pp-waves in 11-dimensions}
J.~P.~Gauntlett and C.~M.~Hull,
``pp-waves in 11-dimensions with extra supersymmetry,''
JHEP {\bf 0206} (2002) 013
[arXiv:hep-th/0203255].
\bibitem{Michelson:A pp-wave}
J.~Michelson,
``A pp-wave with 26 supercharges'',
Class.\ Quant.\ Grav.\ {\bf 19} (2002) 5935-5949,
hep-th/0206204.
\bibitem{LVP:Penrose}
H.~Lu and J.~F.~Vazquez-Poritz,
``Penrose limits of non-standard brane intersections,''
Class.\ Quant.\ Grav.\  {\bf 19} (2002) 4059
[arXiv:hep-th/0204001].
\bibitem{Singh:M5-brane}
H.~Singh,
``M5-branes with 3/8 supersymmetry in pp-wave background,''
Phys.\ Lett.\ B {\bf 543} (2002) 147
[arXiv:hep-th/0205020].
\bibitem{HKS:IIB}
M.~Hatsuda, K.~Kamimura and M.~Sakaguchi,
``From super-AdS$_5\times S^5$ algebra to super-pp-wave algebra,''
Nucl.\ Phys.\ B {\bf 632} (2002) 114
[arXiv:hep-th/0202190].
\bibitem{CHKW:Penrose limit of RG fixed points}
R.~Corrado, N.~Halmagyi, K.~D.~Kennaway and N.~P.~Warner,
``Penrose limits of RG fixed points and pp-waves with background fluxes,''
Adv.\ Theor.\ Math.\ Phys.\  {\bf 6} (2003) 597
[arXiv:hep-th/0205314].
\bibitem{BJLM:penrose limits deformed pp-waves}
D.~Brecher, C.~V.~Johnson, K.~J.~Lovis and R.~C.~Myers,
``Penrose limits, deformed pp-waves and the string duals of $N = 1$
large $N$  gauge theory,''
JHEP {\bf 0210} (2002) 008
[arXiv:hep-th/0206045].
\bibitem{BR:Supergravity}
I.~Bena and R.~Roiban,
``Supergravity pp-wave solutions with 28 and 24 supercharges,''
Phys.\ Rev.\ D {\bf 67} (2003) 125014
[arXiv:hep-th/0206195].
\bibitem{GPS:penrose limit and RG flow}
E.~G.~Gimon, L.~A.~Pando Zayas and J.~Sonnenschein,
``Penrose limits and RG flows,''
JHEP {\bf 0209} (2002) 044
[arXiv:hep-th/0206033].
\bibitem{FP;Maximally}
J.~Figueroa-O'Farrill and G.~Papadopoulos,
``Maximally supersymmetric solutions of ten- and eleven-dimensional
supergravities,''
JHEP {\bf 0303} (2003) 048
[arXiv:hep-th/0211089].
\bibitem{SY:IIA}
K.~Sugiyama and K.~Yoshida,
``Type IIA string and matrix string on pp-wave,''
Nucl.\ Phys.\ B {\bf 644} (2002) 128
[arXiv:hep-th/0208029],\\
%
S.~j.~Hyun and H.~j.~Shin,
``N = (4,4) type IIA string theory on pp-wave background,''
JHEP {\bf 0210} (2002) 070
[arXiv:hep-th/0208074].
\bibitem{Meessen:A small note}
P.~Meessen,
``A small note on pp-wave vacua in 6 and 5 dimensions,''
Phys.\ Rev.\ D {\bf 65} (2002) 087501
[arXiv:hep-th/0111031].
\bibitem{Kowalski-Glikman:Positive}
J.~Kowalski-Glikman,
``Positive energy theorem and vacuum states for the Einstein-Maxwell system,''
Phys.\ Lett.\ B {\bf 150} (1985) 125.
\bibitem{LMO:On d=4 5 6}
E.~Lozano-Tellechea, P.~Meessen and T.~Ortin,
``On $d = 4, 5, 6$ vacua with 8 supercharges,''
Class.\ Quant.\ Grav.\  {\bf 19} (2002) 5921
[arXiv:hep-th/0206200].
\bibitem{BMO}
M.~Blau, P.~Meessen and M.~O'Loughlin,
 ``Goedel, Penrose, anti-Mach: Extra supersymmetries of time-dependent  plane
waves,''
JHEP {\bf 0309} (2003) 072
[arXiv:hep-th/0306161].
\bibitem{Townsend:Killing spinors}
P.~K.~Townsend,
``Killing spinors, supersymmetries and rotating intersecting branes,''
arXiv:hep-th/9901102.
\bibitem{F:On the supersymmetries}
J.~M.~Figueroa-O'Farrill,
``On the supersymmetries of anti de Sitter vacua,''
Class.\ Quant.\ Grav.\  {\bf 16} (1999) 2043
[arXiv:hep-th/9902066].
\bibitem{Ortin:A note}
T.~Ortin,
``A note on Lie-Lorentz derivatives,''
Class.\ Quant.\ Grav.\  {\bf 19} (2002) L143
[arXiv:hep-th/0206159].
\bibitem{S:IIB}
M.~Sakaguchi,
``IIB PP-Waves with Extra Supersymmetries'',
arXiv:hep-th/0306009.
\bibitem{GP;Geometry of D=11}
J.~P.~Gauntlett and S.~Pakis,
``The geometry of D = 11 Killing spinors,''
JHEP {\bf 0304} (2003) 039
[arXiv:hep-th/0212008].
\bibitem{SY}
M.~Sakaguchi and K.~Yoshida,
``M-theory on a time-dependent plane-wave,''
JHEP {\bf 0311} (2003) 030
[arXiv:hep-th/0309025].
\bibitem{HPW}
M.~Blau and M.~O'Loughlin,
``Homogeneous plane waves,''
Nucl.\ Phys.\ B {\bf 654} (2003) 135
[arXiv:hep-th/0212135].
\end{thebibliography}
\end{document}